# Manipulating reflection-type all-dielectric non-local metasurfaces via parity of particle number

Hao Song[1,2], Xuelian Zhang[1,2], Yanming Sun[3], and Guo Ping Wang[3]*

[1]College of Physics and Electronic Information Engineering, Neijiang Normal University, Neijiang, 641112, P. R. China

[2] Lab of Micro-scale Optoelectronic Devices, Neijiang Normal University, Neijiang, 641112, P. R. China

[3]College of Electronics and Information Engineering, Shenzhen University, Shenzhen, 518060, P. R. China

*Email: gpwang@szu.edu.cn

**Abstract**

Parity of particle number is a new degree of freedom for manipulating metasurface, while its influence on controlling non-local metasurfaces remains an unresolved and intriguing question. We propose a metasurface consisting of periodically arranged infinite-long cylinders made from multiple layers of $SiO_2$ and $WS_2$. The cylinder exhibits strong backward scattering due to the overlapping magnetic dipole and electric quadrupole resonances. Without non-local coupling in unit cells, the infinite-size metasurface manifests high reflection across all instances. However, parity-dependent reflectivity diverges with non-local coupling in supercells, exhibiting either increased logarithmic or decreased exponential behavior, with significant distinctions at small particle numbers. Interestingly, equal magnitude reflection and transmission reversals are achievable through alternation between adjacent odd and even particle numbers. The finite-size non-local metasurfaces behave similarly to the infinite-size counterparts, yet high reflection disappears at small particle numbers due to energy leakage. Essentially, high reflection arises from strong backward scattering and effective suppression of lateral multiple scatterings. Our work aids in the actual metasurface design and sheds new light on photonic integrated circuits and on-chip optical communication.

## 1 Introduction

Metasurfaces, an alternative to traditional large-size bulk optical components, are two-dimensional (2D) planar metamaterials consisting of periodically arranged subwavelength scatterers [1-3]. Due to their minimal thickness compared to the incident wavelength, metasurfaces manipulate electromagnetic (EM) waves not through optical path accumulation but by inducing abrupt changes in EM responses between adjacent sides of the scatterers [4]. The primary mechanisms driving this abrupt response are the electric and magnetic resonances in metallic scatterers and the Mie resonances in dielectric scatterers [5, 6]. Consequently, metasurfaces can produce arbitrary chiral and linear polarization responses via gradient phases, such as the Pancharatnan-Berry phase [7,8], and the generalized Snell's law [1, 4, 9, 10]. These features have led to diverse applications, including optical vortex generation [11], subwavelength focusing [12], wavefront deflection [13], holographic imaging [14], and polarization beam splitting [15]. In general, the behavior of local Huygens' metasurfaces is governed by the resonant local modes of each isolated scatterer [16, 17], necessitating sufficient spacing between adjacent scatterers to prevent coupling.

The manipulation mechanisms face significant challenges in metasurfaces with densely packed scatterers. To address these issues, non-local metasurfaces featuring non-local modes between scatterers have emerged [18, 19]. These non-local modes can be generated through two primary pathways [20]: near-field coupling between adjacent scatterers [21, 22] and array lattice involving spatially extended modes such as propagating surface plasmon polaritons [23, 24], lattice resonances [25], and guided-mode resonances [26]. In addition to some constructed theoretical models, the analysis of non-local modes with complex nonlinear couplings is generally carried out through numerical methods, such as the finite element method (FEM), finite-difference time-domain (FDTD), and discrete dipole approximation (DDA) [27, 28]. To date, non-local modes have become a new degree of freedom for manipulating EM waves [27], and non-local metasurfaces have produced numerous exotic effects and practical applications, including bound states in the continuum [29], polaritonic chemistry [30], dynamic wavefront manipulation [31], dark-field edge emission [20], eye tracking [32], and light localization [33], etc.

Although extensive research has explored the manipulating degrees of freedom, such as amplitude, phase, and polarization in both local and non-local metasurfaces, researchers have to face an additional factor in practical finite-size metasurfaces: the particle number of constituent scatterers [34]. Generally, a finite-size metasurface can be fabricated with a sufficient number of scatterers. However, due to the complexity of fabrication, determining the optimal size to truncate the metasurface becomes necessary. Recently, non-local phase gradient metasurfaces (PGMs) have demonstrated beam steering and asymmetric absorption by manipulating the particle number in supercells [35]. Furthermore, controlling the parity of particle number enables the realization of incidence-locking and incidence-free higher-order unidirectional radiation in PGMs [36]. The parity-dependent higher-order diffraction effect has also been observed in non-local optical PGMs with supercells containing fixed particle numbers [37]. However, the comprehensive understanding of parity-dependent reflection and transmission manipulation in non-local all-dielectric metasurfaces remains incomplete.

In this article, we explore the effects of parity on non-local all-dielectric metasurfaces in the near-infrared (NIR) band. Unlike previous studies that employed different unit cells to ensure phase conservation of $2\pi$ within each supercell [35-37] and examined higher-order reflection and transmission responses [36, 37], we utilize identical unit cells and concentrate on the zeroth-order responses. We choose dielectric materials to avoid dissipative losses in the NIR band [6, 38]. Generally, achieving high reflection in dielectric structures requires a high impedance mismatch [39]. However, we designed an infinite-long radial anisotropic cylinder with strong backward scattering to anticipate obtaining high reflectivity. The cylinder is composited by multiple-layer (ML) $SiO_2$ and $WS_2$. First, we discuss the reflection of an infinite-size local metasurface composed of periodically arranged cylinders. We then investigate the reflection of infinite-size non-local metasurfaces with supercells containing odd or even cylinders. The reflectivity relies on not only the spacing but also the parity. Next, we explore the finite-size non-local metasurfaces. Finally, we discuss the physical mechanisms behind high reflection via multiple scatterings. Our work contributes to deciding on the design size for the practical metasurface and sheds new light on nonlinear photonic integrated circuits and on-chip optical communication.

## 2 Results and discussions
### 2.1 Strong backward scattering of anisotropic cylinder

We begin by investigating the scattering properties of an isolated infinite-long SiO$_2$ cylinder with a refractive index $n_l$ of 1.45 [40]. The low-index material is advantageous to avoid several limitations of high-index materials such as difficulty in generating complex patterns, performance degradation post-fabrication, and inhibiting efficient coupling in photonic circuits [41-44]. Additionally, SiO$_2$ has been applied in various scenarios such as radiative cooling [45], drug delivery [46], optoelectronic devices [47], lithium-ion batteries [48], etc. A TE plane EM wave normally incident, characterized by magnetic field **H** paralleling on the center axis of the cylinder, as the schematic shown in Figure 1(a). Here, the incident wavelength ($\lambda$) is 1550 *nm* and the wavevector is **k**. According to the Mie scattering theory [49], the *m*th-order scattering coefficient can be written as

$$a_m = \frac{n_l J'_m(x_r) J_m(n_l x_r) - J_m(x_r) J'_m(n_l x_r)}{n_l J_m(n_l x_r) H_m^{(1)'}(x_r) - J'_m(n_l x_r) H_m^{(1)}(x_r)} \quad . \tag{1}$$

Here, $J_m(x)$ and $H_m^{(1)}(x)$ are the first kind of Bessel and Hankel functions respectively, and the size parameter $x_r$ equals wavenumber $k$ times the radius $r$. Then, the scattering efficiency normalized by single-channel scattering limit $2\lambda/\pi$ [50, 51] is expressed by

$$N_{sca} = |a_0|^2 + 2\sum_{m=1}^{\infty} |a_m|^2 \quad . \tag{2}$$

Figure 1(b) shows the scattering efficiency (blue curve) of the cylinder based on the scattering theory (ST). The curve increases monotonically with increasing $x_r$, indicating that the cylinder does not excite any significant scattering resonant peaks. To gain deeper insight into the Mie modes of the cylinder, the multipolar scattering spectra are shown in Figure S1(a) [see Supplement 1]. The coexistence of magnetic dipole (MD), electric dipole (ED), electric quadrupole (EQ), and electric octupole (EO) modes results in the total scattering (red curve), namely the blue curve in Figure 1(b). Due to the rotational symmetry of cylinder coordination, MD mode has only one scattering channel but electric multipoles possess two [49, 52, 53]. Thus, the maximum normalized efficiencies of resonant MD and electric multipoles correspond to 1 and 2. Accordingly, Figure S1(a) indicates that the cylinder does not possess resonant multipoles. To verify the analysis results, the scattering efficiency is numerically simulated using commercial software COMSOL based on the FEM [54]. Perfectly matched layers (PMLs) are implemented around the cylinder to truncate the infinite air space. Consequently, the scattering (red circles) coincides precisely with the curve. Therefore, it confirms the validity of the ST analysis and the feature of the absence of resonant peaks.

However, high-reflection dielectric metasurfaces usually need scatterers with resonant modes [52, 55-57]. We propose an ML radial anisotropic cylinder by ML WS$_2$ and SiO$_2$ to improve the scattering. For an incident wavelength of 1550 nm, WS$_2$ demonstrates anisotropic refractive indices with a radial refractive index $n_r$ of 2.447 and an azimuthal refractive index $n_t$ of 3.756 [58]. Figure 1(c) demonstrates the schematic and the same TE wave illuminates the cylinder. The WS$_2$ and SiO$_2$ layers are alternatively arranged 15 times. Define filling factor $f_1$ as the thickness ratio of the total WS$_2$ to the ML cylinder. Here, $f_1$ is 0.695. Then, the ML cylinder can be effectively replaced by a homogenous cylinder with effective material parameters when the thickness of each layer far less than the incident wavelength. Consequently, based on the effective medium theory (EMT) [53], the anisotropic radial ($n_{rh}$) and azimuthal ($n_{th}$) refractive indices are expressed as

$$n_{th} = \sqrt{(1-f_1)n_l^2 + f_1 n_t^2} \quad , \tag{3}$$

$$n_{rh} = n_l n_r \Big/ \sqrt{f_1 n_l^2 + (1-f_1) n_r^2} \quad . \tag{4}$$

The anisotropic parameter $\eta$ is defined as $\eta = n_{th} / n_{rh}$. For the ML cylinder, $\eta$ is 1.652. According to the Mie theory, the scattering coefficients of the radial anisotropic cylinder are written as [53, 58]

$$b_m = b_{-m} = \frac{n_{th} J_{\tilde{m}}(n_{th} x_r) J'_m(x_r) - J_m(x_r) J'_{\tilde{m}}(n_{th} x_r)}{n_{th} J_{\tilde{m}}(n_{th} x_r) H^{(1)'}_m(x_r) - H^{(1)}_m(x_r) J'_{\tilde{m}}(n_{th} x_r)} , \tag{5}$$

where $\tilde{m}$ is the radial anisotropy-revised function order $\tilde{m} = m\eta$. As a result, the normalized scattering efficiency is similar to Eq. 2, but the $a_0$ is substituted by $b_0$ and $a_m$ is interchanged with $b_m$, respectively.

Subsequently, Figure 1(d) shows the ML cylinder's scattering efficiency (ST, blue curve). There are three significant resonant peaks at $x_r$=1.381, 1.972, and 2.544, respectively. Thus, the resonant modes have emerged. To verify the result, we simulate the scattering efficiencies of the ML cylinder with effective material parameters (eff-sim) and the practical anisotropic ML cylinder (sim), respectively. Their numerical results are marked by the green curve and the red circles, respectively. The green curve and circles agree well with the blue curve, which demonstrates the validity of the scattering theory analysis. Therefore, the ML cylinder exhibits strong resonances at the three points.

Then, Figure 2(a) shows the normalized scattering efficiency of multipoles of the ML cylinder. Specifically, the ML cylinder excites the resonant ED mode and weak MD mode at the first peak, the MD and EQ resonances at the second peak, and the EO resonance as well as strong ED and MD modes at the last peak. Generally speaking, strong backward scattering of scatterers contributes to achieving high reflection in local metasurfaces. Thus, we focus on the ML cylinder with the strong backward scattering. Ideally, independent resonant dipoles behave the strong backward scattering, whereas multiple modes are often excited simultaneously in an actual dielectric cylinder. To date, the physical understanding of mode interferences inducing strong backward scattering is not yet complete, a common method is adopting the second kind of Kerker condition originating in the interference of anti-phase ED and MD modes [59, 60]. Fortunately, the strong backward scattering can be achieved at the second peak here. For convenience, point A denotes the second peak, i.e. $x_r$=1.971. Therefore, we will concentrate on point A.

Based on the accessible multipolar scattering coefficients, angular scattering intensity is expressed as [49, 53]

$$I_{SA}(\theta) = \frac{2}{\pi k} \left| b_0 + 2\sum_{m=1}^{\infty} b_m \cos(m\theta) \right|^2 , \tag{6}$$

where $\theta$ is the scattering angle, and the backward direction corresponds to the 180 degrees. Figure 2(b) demonstrates the angular scattering spectra of the ML cylinder at point A. The interference between the resonant MD (blue curve) and EQ (black curve) modes results in the total angular scattering (red curve), yet the weak ED and EO modes hardly contribute to total scattering. Importantly, the red curve suggests the emergence of strong backward scattering. Meanwhile, the scattering intensity ratio of the lateral (at 90 or 270 degrees) to backward is 0.114, which indicates the lateral scatterings are very weak. On the other hand, to quantify the directionality, an asymmetry parameter is defined as

$$g = \frac{\oint_s \cos\theta I_{SA}(\theta) ds}{\oint_s I_{SA}(\theta) ds} , \tag{7}$$

where *s* is a surface enclosing the particle. In particular, *g*=0 is the symmetry light scattering, while the *g* is closer to ±1, the directionality is better. Here, the *g* is 0.018 approaching 0, evidencing the symmetric scattering at point A. Consequently, the forward (0 degree) scattering is nearly as strong as the backward, and the profile is similar to a dipole. Therefore, the constructive interference between the MD and EQ resonances enhances backward scatterings, whereas their destructive interference suppresses the lateral scatterings. It is worth mentioning that the strongest backward scattering occurs at point A, and the forward scattering cannot be completely suppressed due to the incident wave [60, 61].

To verify the angular scattering spectra, the scattering electric field |$\mathbf{E}_s$| distribution of the ML cylinder with effective material parameters is depicted in Figure 2(c). One can see that the strongest electric field occurs inside the cylinder and the outside field is predominantly in forward and backward, which are consistent with the angular scattering spectra. Moreover, Figure 2(d) shows the |$\mathbf{E}_s$| distribution of the actual ML cylinder based on the FEM. The |$\mathbf{E}_s$| distribution also dominant in forward and backward indicates the validity of the EMT method. For simplicity, we will focus on the effective material parameter model in the following.

For comparison, Figure 2(e) shows the angular scattering spectra of an isolated pure $SiO_2$ cylinder at point A. According to Figure S1(a), the cylinder excites off-resonant MD (the blue), ED (the green), and weak EQ (the black) modes at point A. The interferences of these modes lead to the total angular scattering spectrum (the red) as indicated in Figure 2(e). The *g*=0.775. Consequently, the cylinder exhibits significant forward scattering but without backward scattering. This scattering feature is verified by the simulated |$\mathbf{E}_s$| distribution in Figure S1(b). Therefore, the ML cylinder enables significantly enhanced backward scattering compared to the pure $SiO_2$ one at point A.

**2.2 Infinite-size local metasurface**

Figure 3(a) shows a schematic of an infinite-size metasurface composed of periodically arranged ML cylinders at point A. *dP* is the minimum surface spacing between adjacent cylinders and is normalized by the λ. Consequently, the metasurface only has upward and downward radiation channels due to its opening in the *y* direction [62, 63]. Notice that, here, just the zeroth-order upward reflectivity (*R*) and downward transmitivity (*T*) responses are considered when the same TE wave normally illuminates the metasurface.

Next, we achieve the *R* and *T* based on the FEM. The above results have displayed the local modes of the ML cylinder, i.e. the localized Mie resonances. Without considering non-local couplings between adjacent cylinders, a unit cell containing one ML cylinder is adopted for the calculation. The periodic boundary condition (PC) is performed on both edges of the unit cell perpendicular to the *x*-direction as shown in the inset of Figure 3(a). Distance between the two PCs is the period (*P*) equal to *dP*+2*r*. Furthermore, the Floquent periodicity induces a phase difference of $e^{-k_xP}$ between the two PCs, and the Bloch wave will emerge due to the discrete translational symmetry in the *x* direction [64]. Yet, the two PCs here are identical since the $k_x$=0. On the other hand, the PMLs are applied to truncate the infinite air domain in the *y* direction. As a result, the *R* and *T* merely depend on the interferences of local modes of all cylinders [52] and they are calculated by the input and output energy from the discontinuous top and bottom interfaces of the metasurface.

For the sake of consistency, we adopt energy flux to define the zeroth-order *R* and *T* under normal incidence throughout this paper, thus, written as,

$$R = \int_l \frac{p_r}{p_i} dl \ , \ T = \int_l \frac{p_t}{p_i} dl \ . \tag{8}$$

Here, the $p_r$ and $p_t$ are the outflow time-average energy densities from the top and bottom interfaces [65], respectively,

$$p_{r,t} = \left| \frac{1}{2} \text{Re}(\mathbf{E}_{r,t} \times \mathbf{H}^*_{r,t}) \right| . \tag{9}$$

$l$ denotes the length of input and output ports. The $p_i$ is the incident energy density is calculated as

$$p_i = \frac{1}{2} c\varepsilon |\mathbf{E}|^2 , \tag{10}$$

$c$ is the light speed, $\varepsilon$ is the dielectric constant of the environment. According to the energy conservation, the responses of this infinite-size metasurface fulfill $R+T=1$.

Figure 3(b) depicts the $R$ (the red) and $T$ (the blue) of the local metasurface as a function of $dP$. To ensure the zeroth-order diffraction, $dP$ changes within the range from 0 to 0.3726, derived from the grating equation [66]. One can see that $R$ rises sharply as $dP$ increases from 0 to 0.06, then becomes nearly stable and its value exceeds 0.9. Figures 3(c)-3(e) show the electric field $|\mathbf{E}_x|$ distributions of one unit cell at $dP$=0.1, 0.2, and 0.3, respectively. The significant standing wave fringes originating from the interference between incident and reflected electric field occur in the reflection space, whereas nearly no $|\mathbf{E}_x|$ distribution in the transmission domain. Thus, the local metasurface composed of the strong backward scattering cylinders is a high reflection in the range.

## 2.3 Infinite-size non-local metasurface

Previous works [67, 68] suggest that the non-local couplings between adjacent Mie resonant scatterers cannot be neglected, significantly the higher-order multipolar resonances cases. In our work, pronounced near-field couplings between adjacent cylinders occur due to the resonant MD and EQ modes. Thus, this section will investigate the $R$ and $T$ of the infinite-size metasurface with non-local couplings. We propose a supercell composed of $np$ identical cylinders as depicted in Figure 4(a). The PCs are applied to the left and right edges of the supercell while open in the top and bottom directions. Here, the $l=np \times P$. Therefore, the supercell not only has local modes but also non-local couplings.

Figure 4(b) shows the $R$ as a function of $dP$ and odd $np$. The $np$ changes from 1 to 51. To simplify, we define three $R$ regions, namely the high reflection region ($R \geq 0.8$), the transition region ($0.8 > R > 0.2$), and the low reflection region ($R \leq 0.2$). The dashed blue, white, and red lines denote the critical boundaries of the regions, respectively. For instance, a point ($dP$, $np$) makes $R \geq 0.8$ on the left side of the blue line, while no such point is on the right. This definition can be extended correspondingly to the white and red lines. Hence, the metasurface behaves as a high reflection within wide ranges of $dP$ and $np$ on the left of the blue line. As for the other three regions, the $R$ depends on the two parameters. High transmission with $T>0.93$ occurs in the low reflection region between the white and red lines, more details are depicted in Figure S2(a). For $np \leq 3$, the metasurface maintains a high reflection throughout the range. On the other hand, Figure 4(c) describes the $R$ map concerning $dP$ and even $np$. The $np$ varies from 2 to 50. On the left side of the blue line, the metasurface maintains a high reflection regardless of $dP$ and $np$. In contrast, the $R$ is sensitive to $dP$ and $np$ in the other three regions. Additionally, the maximum $T$ is as high as 0.97 in the low reflection region, more details can be seen in Figure S2(b).

To gain more insights into the parity, we investigate the $R$ and $T$ as a function of $np$ when $dP$ is fixed. Four points B-E from the four regions are randomly selected, namely $dP$=0.1104, 0.236, 0.295, and 0.3725, respectively. The odd $np$ is from 1 to 59, the even $np$ is from 2 to 60. For point B, Figure 5(a) shows the $R$ of the odd $np$ (Rodd, the red square) and the even $np$ (Reven, the red circle), and the $T$ of the odd $np$ (Todd, the blue square) and the even $np$ (Teven, the blue circle), respectively. Specifically, the metasurface exhibits high reflection in all *np*. Due to the infinite periodicity, the Rodd+Todd=1 and Reven+Teven=1. Rodd and Reven have the same trend, *R* rapidly increases and then gradually stabilizes and approaches total reflection. A general mathematical model well fits the trend, expressed as,

$$F(x) = a_1 \exp(b_1 x) + a_2 \exp(b_2 x), \quad (11)$$

where *x* denotes the *np*. $a_1$, $b_1$, $a_2$, and $b_2$ are variable coefficients. The detailed values of the four parameters at points B-E can be found in the S3 of Supplement 1. It is worth noting that several abrupt data have been excluded during fitting, e.g. *np*=1 and 2. Indeed, the trend can also be equivalently described as an increased logarithmic-like function, i.e. $\ln(x)$. Moreover, Reven is greater than its adjacent Rodd. Assume *dR* to describe the difference between Reven and its adjacent Rodd, then *dT* denotes the case of *T*. The *dR* is more pronounced at smaller *np*, while Rodd gradually approaches Reven at larger *np*. Therefore, the parity effect is significant at small *np*. On the other hand, the trends of *T* curves satisfy a decayed exponential-like function. The *dT* is also remarkable at smaller *np* but approaches zero at larger *np*.

Next, Figure 5(b) depicts the *R* and *T* as functions of *np* at point C in the transition region I between the dashed blue and white lines. Rodd and Todd maintain trends similar to the increased logarithmic-like and decayed exponential-like functions, respectively. On the contrary, the Reven and Teven separately reverse to the decayed exponential-like and increased logarithmic-like functions. For *np*≤5, Rodd and Todd are away from the trends. Similarly, the *dR* and *dT* are pronounced at smaller *np*, gradually decreasing to zero at larger *np*. Then, Reven is greater than its adjacent Rodd. But, the stable *R* is no longer a high reflection at larger *np*.

Continuously, the *R* and *T* at point D in the low reflection region are demonstrated in Figure 5(c). Differently, Rodd and Reven behave as decayed exponential-like functions, while Todd and Teven manifest as increased logarithmic-like functions. Reven decreases to 0 much quicker resulting in a total transmission (Teven=1). Hence, Rodd is greater than its adjacent Reven. *dR* and *dT* are also evidently at smaller *np* while can be neglected at larger *np*. Finally, Figure 5(d) shows the *R* and *T* at point E in the transition region II beyond the red line. Rodd and Todd exhibit the decayed exponential-like and increased logarithmic-like functions, respectively. However, Reven~0.55 and Teven~0.45 are almost unchanged in the range. In the same way, the trens of *dR* and *dT* are the same as above.

Interestingly, besides the high reflection region, the high reflection can also be achieved in other regions at small *np*, for example, at *np*=1, 2, and 3 in the transition region I, *np*=1 and 3 in the low reflection region, and *np*=1 and 3 in the transition region II. Additionally, in the transition region I, *T* =0.674 translates to *R* =0.705 when *np* varies from 7 to 8, as depicted in the black dashed elliptical box of Figure 5(b). Figure 6(a) shows the |$\mathbf{E}_x$| distribution of one supercell at *np*=8. The significant standing wave fringes indicant the emergence of high reflection. Figure 6(b) describes the $\mathbf{E}_x$ at *np*=7. The transmitted $\mathbf{E}_x$ suggests the generation of high transmission. Similarly, in the low reflection region, high reflection (*R*=0.874) at *np*=3 turns into high transmission (*T*=0.892) at *np*=4 as shown in the box of Figure 5(c). The |$\mathbf{E}_x$| at *np*=3 in Figure 6(c) and $\mathbf{E}_x$ at *np*=4 in Figure 6(d) validate the high reflection and the high transmission, respectively. Realizing a nearly equal magnitude reversal between high reflection and high transmission via adding or removing one particle in the supercell offers an efficient optical switch.

**2.4 Finite-size non-local metasurface**

In practical applications, it is impossible to fabricate a perfect infinite-size periodic structure. In this sense, we focus on the truncated finite-size metasurface in the following. Figure 7(a) presents the schematic of a finite-size metasurface consisting of *np* periodically arranged ML cylinders. Compared to the infinite-size one, the PCs have changed to the PMLs. The *l*=*np*×*P*. Rodd and Reven as functions of *dP* and *np* are plotted in Figures 7(b) and 7(c), respectively. Aligned with the infinite-size structure, the critical lines are marked out by the same three dashed lines. The *R* has the same manner as the infinite-size case at large *np*, yet different at small *np*. In addition, Todd and Teven can be seen in Figures S3(a) and S3(b), correspondingly.

Then, *R* and *T* as functions of *np* at point B are shown in Figure 8(a). The detailed fitting parameters can be found in the S5 of Supplement 1. Rodd and Reven alter synchronously obeying an increased logarithmic-like function, but *np*=3 is an abrupt data. Reven is slightly greater than its adjacent Rodd. The high reflection is achieved at a larger *np*, whereas all *T* is the low transmission. *R*+*T*≠1 originates from the energy radiation leakage on the open lateral boundaries. The smaller the *np*, the more energy loss. Moreover, smaller *np* means higher relative interface roughness, leading to radiation losses in other non-normal directions. A general way to enable the response of the finite-size to approach the infinite-size one is adding cylinders since more lateral leaky energy will be suppressed [34].

Figure 8(b) displays the *R* and *T* at point C. The responses sharply change when *np*<20 but maintain stability (*R*~0.5, *T*~0.43) at larger *np*. Then, Figure 8(c) demonstrates the *R* and *T* at point D. The *R* responses satisfy the decayed exponential-like functions, and Reven decays faster. Rodd is larger than its adjacent Reven, while Teven is larger than its adjacent Todd. Finally, the *R* and *T* at point E are shown in Figure 8(d). Reven, Teven, and Todd fit the increased logarithmic-like functions, whereas Rodd acts like a decayed exponential-like function. The stable *R* and *T* are close to 0.45 and 0.41, respectively.

Furthermore, the nearly equal magnitude response reversal also be found here. The dominant reflection with *R*=0.494 at *np*=12 turns into the dominant transmission with *T*=0.498 at *np*=13, illustrated in the black dashed elliptical box of Figure 8(b). The |$\mathbf{E}_x$| distribution in Figure 9(a) verifies the dominant reflection, and the $\mathbf{E}_x$ in Figure 9(b) confirms the dominant transmission. Similarly, the dominant transmission with *T*=0.436 at *np*=8 switches to the dominant reflection with *R*=0.438 at *np*=9, marked by the box in Figure 8(c). The |$\mathbf{E}_x$| in Figure 9(c) and the $\mathbf{E}_x$ in Figure 9(d) validate the dominant reflection and transmission, respectively.

Next, considering the imperfect fabrication, the disordered positions of the ML cylinders are introduced in the high-reflection finite-size metasurface. According to Figures 7(b) and 7(c), each *dP* is randomly located in the range of 0.01 to 0.183. Figure 10(a) illustrates the *R* of the random metasurface composed of odd *np* ML cylinders. The red circle denotes the *R* of the case of *np*=35 and the red line is the average *R* of 10 samples. The average value and standard deviation are about 0.804 and 0.0152, respectively. Hence, the metasurfaces are high reflection. Then, for *np*=45, the average value and standard deviation are about 0.833 and 0.016, respectively. Compared to the *np*=35, the *R* increases, and the standard deviation decreases. Furthermore, for *np*=55, the average value and standard deviation are about 0.839 and 0.011, respectively. The *R* and standard deviation trend to stable.

Then, the *R* of even *np* cylinders case is shown in Figure 10(b). For *np*=34, the average value and standard deviation are 0.805 and 0.0171, respectively. For *np*=44, the average is 0.831 and the standard deviation is 0.0170. For *np*=54, the average is 0.838 and the standard deviation is 0.0089. To sum up, all random metasurfaces have high reflection, and the *R* increases and gradually becomes stable as *np* increases. Therefore, regardless of the parity, the metasurface with a larger *np* exhibits good robustness against the positional disorder.

## 2.5 Multiple scatterings

To insight into the physical mechanism, we explore the multiple scatterings relating to the parity. The odd *np* varies from 1 to 11. The scattering calculations are implemented based on the FEM method. Figure 11(a) indicates the angular scattering amplitude spectra at point B. Here, the lateral scattering is defined as the non-forward and non-backward scatterings. The amplitude in the forward and backward are approximately proportional to *np*. New lateral scattering peaks emerge as *np* increases. However, the lateral scattering amplitude is very weak contrary to the forward and backward. Subsequently, Figures 11(b) - 11(d) depict the angular spectra at points C - E, respectively. Their forward and backward amplitudes are unequal and no longer proportional to *np*, beside the lateral scatterings become significant and cannot be neglected. Especially, in Figure 11(b), after *np*=7, the amplitude in the forward and backward no longer increases, while the lateral rapidly rises.

Next, for even *np* changing from 2 to 10, the angular spectra of the four points are shown in Figures 12(a)-12(d). Here, the case of *np*=1 is shown as a comparison. At point B, the amplitudes in the forward and backward increase proportionally to *np*, and the lateral scatterings can be ignorance. Conversely, at point D in Figure 12(c), the forward and backward amplitudes are unequal, and the lateral are significant. Furthermore, the angular spectra of a two-cylinder system are demonstrated in Figure S4. The smaller the *dP*, the stronger the backward scattering, but both the backward and lateral scattering changes are not significant.

For quantitative analysis, parameter $\alpha$ is defined as the backward multiple scattering intensity normalized by the intensity of the individual cylinder. Assume a deviation tolerance of ±10 degrees for all directions in practical application, thus the $\alpha$ is written as

$$\alpha = \frac{\int_{170}^{190} |\mathbf{E}_s(\theta)|^2_{np} \, d\theta}{\int_{170}^{190} |\mathbf{E}_s(\theta)|^2_{1} \, d\theta} \quad , \tag{12}$$

where $|\mathbf{E}_s(\theta)|_{np}$ is the multiple scattering amplitude of *np* cylinders at the $\theta$ direction. Consequently, the $\alpha$ of odd *np* cylinders is depicted in Figure 13(a). The values of red, green, blue, and black circles are calculated based on Figures 11(a)-11(d), respectively. When *np*≤5, all the circles maintain small values but the red is the smallest, and the corresponding finite-size metasurfaces are the low reflection as the Rodd shown in Figures 8(a)-8(d). However, when *np*≥7, the red rapidly rises and becomes the largest, and the corresponding Rodd sharp increases as depicted in Figure 8(a). Therefore, the backward multiple scatterings are directly positively associated with reflectivity.

Furthermore, suppose a parameter $\beta$ to denote the lateral multiple scattering intensity ratio of the *np* cylinders to the single one, namely

$$\beta = \frac{\int_{10}^{170} |\mathbf{E}_s(\theta)|^2_{np} \, d\theta + \int_{190}^{350} |\mathbf{E}_s(\theta)|^2_{np} \, d\theta}{\int_{10}^{170} |\mathbf{E}_s(\theta)|^2_{1} \, d\theta + \int_{190}^{350} |\mathbf{E}_s(\theta)|^2_{1} \, d\theta} \quad . \tag{13}$$

Then, the $\beta$ of odd *np* cylinders is shown in Figure 13(b). When *np*≤5, the values of all circles are small, indicating the lateral energy radiation is less. However, the green, blue, and black circles grow near-linearly as increasing *np*, whereas the red remains small. As a result, the metasurfaces at points C - E radiate massive lateral energy but very little at point B. Therefore, the large Rodd of the metasurface at point B originates in the strong backward multiple scatterings and well-suppressed lateral multiple scatterings.

Then, the $\alpha$ of even $np$ cylinders is depicted in Figure 13(c). For 2 cylinders, the values of all circles are small leading to low reflection. Yet, the red increases near-linearly and behaves as the largest. Thus the strongest backward multiple scatterings occur at point B. Subsequently, the $\beta$ of even $np$ cylinders is shown in Figure 13(d). At $np$=2, all circles demonstrate nearly the same small value. More details of $\alpha$ and $\beta$ of the two-cylinder system concerning $dP$ can be seen in Figures S5(a) and S5(b). The red maintains the smallest as $np$ increases, while the other color circles show different degrees of growth. Thus, the minimum lateral energy leak emerges at point B. The results agree well with the Reven in Figures 8(a)-8(d).

Overall, the $R$ and $T$ have significant parity-dependent due to the parity-dependent $\alpha$ and $\beta$. However, the high reflection at point B essentially originates from the strong backward and well-suppressed lateral multiple scatterings. It is also suitable for the infinite-size metasurfaces. Moreover, the significant lateral multiple scatterings enable us to explain the R+T≠1 for the finite-size metasurface.

**Conclusion**

This article explores the NIR parity-dependent high reflection in the all-dielectric non-local metasurface. We design an ML radial anisotropic cylinder composed of alternating layers of $SiO_2$ and $WS_2$. This cylinder induces significant backward scattering due to the overlap of MD and EQ resonances. The consistency between analytical and numerical results suggests that the cylinder can be accurately modeled using effective material parameters based on the EMT. We begin with the infinite-size local metasurface, which demonstrates high reflectivity across the entire spacing range. Then, when considering non-local couplings within the supercell, both the odd and even configurations demonstrate high reflection at small spacings. However, only the odd small particle number achieves high reflection at large spacings. For small spacings, the reflectivities of both the odd and even increase logarithmically. However, at large spacings, the odd and the even have different manners, i.e. the increased logarithmic-like or the decayed exponential-like function. The disparity in parity-dependent reflectivity is pronounced at the small particle number but diminishes at the large. Interestingly, a reversal between equal amplitude reflection and transmission is realized by manipulating the adjacent odd and even particle numbers. Furthermore, the high reflection of the finite-size non-local metasurface is similar to those of the infinite-size but disappears at the small particle number due to the lateral energy leakage.

The primary mechanism behind the observed high reflection is the combination of strong backward and well-suppressed lateral multiple scatterings. However, the scatterings also rely on the parity. Furthermore, the observed reversal is distinct from previously reported cases involving changes in spacing between meta-atoms [69] or alterations in the abrupt phase of meta-atoms in phase gradient metasurfaces [37]. It is important to note that while our work focuses on non-local couplings within the supercell, non-local lattice resonances [29, 63] may also be present in our local metasurface. Future investigations will explore active metasurfaces capable of controlled radiation through parity manipulation, offering an alternative to complex encoding techniques [70, 71]. Our work verifies the new degree of freedom of the parity, facilitates size optimization for practical finite-size metasurfaces, and opens new prospects for nonlinear on-chip communication and integrated photonic circuits.

**Acknowledgments**

This work was supported by the National Natural Science Foundation of China (12304425, 12074267); Sichuan Science and Technology Program (2024NSFSC1354); National Key Research and Development Program of China (2022YFA1404500); Basic Research and Applied Basic Research Project of Neijiang City (NJJH202302).

**Disclosures**

The authors declare no competing financial interest.

**Data availability.** Data may be obtained from the authors upon reasonable request.

**Supplemental document.** See Supplement 1 for supporting content.

# References

[1] N. Yu, P. Genevet, M. A. Kats, F. Aieta, J.-P. Tetienne, F. Capasso, Z. Gaburro, Light propagation with phase discontinuities: generalized laws of reflection and refraction, Science, 334(6054), 333-337 (2011).

[2] S. Sun, K. Y. Yang, C. M. Wang, T. K. Juan, W. T. Chen, C. Y. Liao, Q. He, S. Xiao, W.-T. Kung, G.-Y. Guo, L. Zhou, D. P. Tsai, High-efficiency broadband anomalous reflection by gradient meta-surfaces, Nano Lett., 12(12), 6223-6229 (2012).

[3] S. Sun, Q. He, S. Xiao, Q. Xu, X. Li, and L. Zhou, Gradient-index meta-surfaces as a bridge linking propagating waves and surface waves, Nat. Mater. 11(5), 426-431 (2012).

[4] Q. He, S. Sun, S. Xiao, L. Zhou, High-efficiency metasurfaces: principles, realizations, and applications, Adv. Optical Mater. 6(19), 1800415 (2018).

[5] H. H. Hsiao, C. H. Chu, D. P. Tsai, Fundamentals and applications of metasurfaces, Small Methods, 1(4), 1600064 (2017).

[6] S. Sun, Q. He, J. Hao, S. Xiao, L. Zhou, Electromagnetic metasurfaces: physics and applications, Adv. Opt. Photon. 11(2), 380-479 (2019).

[7] E. Cohen, H. Larocque, F. Bouchard, F. Nejadsattari, Y. Gefen, E. Karimi, Geometric phase from Aharonov-Bohm to Pancharatnam-Berry and beyond, Nat. Rev. Phys., 1, 437-449 (2019).

[8] X. Xie, M. Pu, J. Jin, M. Xu, Y. Guo, X. Li, P. Gao, X. Ma, X. Luo, Generalized Pancharatnam-Berry phase in rotationally symmetric meta-atoms, Phys. Rev. Lett. 126(18), 183902 (2021).

[9] S. Shen, Z. Ruan, Y. Yuan, H. Tan, Conditions for establishing the "generalized Snell's law of refraction" in all-dielectric metasurfaces: theoretical bases for design of high-efficiency beam deflection metasurfaces, Nanophotonics, 11(1), 21-32 (2022).

[10] L. Li, Y. Shi, T. J. Cui, Electromagnetic Metamaterials and Metasurfaces: From Theory To Applications, Springer, Singapore, (2024).

[11] D. Wang, F. Liu, T. Liu, S. Sun, Q. He, L. Zhou, Efficient generation of complex vectorial optical fields with metasurfaces, Light Sci. Appl., 10, 67 (2021).

[12] X. Lu, Y. Guo, M. Pu, Y. Zhang, Z. Li, X. Li, X. Ma, X. Luo, Broadband achromatic metasurfaces for sub-diffraction focusing in the visible, Opt. Express, 29(4), 5947-5958 (2021).

[13] X. Fu, F. Yang, C. Liu, X. Wu, T. J. Cui, Terahertz beam steering technologies: from phased arrays to field-programmable metasurfaces, Adv. Optical Mater., 8(3), 1900628 (2019).

[14] H. Ren, G. Briere, X. Fang, P. Ni, R. Sawant, S. Héron, S. Chenot, S. Vézian, B. Damilano, V. Brändli, S. A. Maier, P. Genevet, Metasurface orbital angular momentum holography, Nat. Commun. 10, 2986 (2019).

[15] S. Kang, B. Zhou, Y. Xie, J. Wang, W. Jia, C. Zhou, Polarization beam splitter based on 2D transmissive grating, Opt. Express 32(12), 20589-20599 (2024).

[16] N. Meinzer, W. L. Barnes, I. R. Hooper, Plasmonic meta-atoms and metasurfaces, Nat. Photonics, 8, 889-898 (2014).

[17] K. Koshelev, Y. Kivshar, Dielectric resonant metaphotonics, ACS Photonics, 8(1), 102-112 (2021).

[18] K. Shastri, F. Monticone, Nonlocal flat optics, Nat. Photonics, 17, 36-47 (2023).

[19] J. Yao, R. Lin, M. K. Chen, D. P. Tsai, Integrated-resonant metadevices: A review, Adv. Photonics, 5(2), 024001 (2023).

[20] J. Yao, W.-L. Hsu, Y. Liang, R. Lin, M. K. Chen, D. P. Tsai, Nonlocal metasurface for dark-field edge emission, Sci. Adv. 10(16), eadn2752 (2024).

[21] Q. Zhao, H. Zhang, Z.-K. Zhou, X.-H. Wang, Enhancing chiroptical responses in the nanoparticle system by manipulating the far-field and near-field couplings, Opt. Express, 31(6), 9376-9386 (2023).

[22] R. Chai, Q. Liu, W. Liu, Z. Li, H. Cheng, J. Tian, S. Chen, Emerging planar nanostructures involving both local

and nonlocal modes, ACS Photonics, 10(7), 2031-2044 (2023).

[23] Q. Xu, Y. Lang, X. Jiang, X. Yuan, Y. Xu, J. Gu, Z. Tian, C. Ouyang, X. Zhang, J. Han, Meta-optics inspired surface plasmon devices, Photonics Insights, 2(1), R02 (2023).

[24] K. V. Sreekanth, J. Perumal, U. Dinish, P. Prabhathan, Y. Liu, R. Singh, M. Olivo, J. Teng, Tunable Tamm plasmon cavity as a scalable biosensing platform for surface enhanced resonance Raman spectroscopy, Nat. Commun. 14, 7085 (2023).

[25] M. S. Bin-Alam, O. Reshef, Y. Mamchur, M. Z. Alam, G. Carlow, J. Upham, B. T. Sullivan, J.-M. Ménard, M. J. Huttunen, R. W. Boyd, Ultra-high-Q resonances in plasmonic metasurfaces, Nat. Commun. 12, 974 (2021).

[26] L. Huang, R. Jin, C. Zhou, G. Li, L. Xu, A. Overvig, F. Deng, X. Chen, W. Lu, A. Alù, Ultrahigh-Q guided mode resonances in an All-dielectric metasurface, Nat. Commun. 14, 3433 (2023).

[27] J. Lin, Q. Li, M. Qiu, Q. He, L. Zhou, Coupling between Meta-atoms: a new degree of freedom in metasurfaces manipulating electromagnetic waves, Chinese Optics, 14(4), 717-735 (2021).

[28] C. Damgaard-Carstensen, S. I. Bozhevolnyi, Nonlocal electro-optic metasurfaces for free-space light modulation, Nanophotonics, 12(14), 2953-2962 (2023).

[29] Y. Tang, Y. Liang, J. Yao, M. K. Chen, S. Lin, Z. Wang, J. Zhang, X. G. Huang, C. Yu, D. P. Tsai, Chiral bound states in the continuum in plasmonic metasurfaces, Laser Photonics Rev., 17(4), 2200597 (2023).

[30] F. Verdelli, Y.-C. Wei, K. Joseph, M. S. Abdelkhalik, G. Masoumeh, S. H. C. Askes, A. Baldi, E. W. Meijer, J. G. Rivas, Polaritonic chemistry enabled by non-local metasurfaces, arXiv:2402.15296.

[31] T. Liu, D. Zhang, W. Liu, T. Yu, F. Wu, S. Xiao, L. Huang, A. E. Miroshnichenko, Phase-change nonlocal metasurfaces for dynamic wave-front manipulation, Phys. Rev. Applied, 21(4), 044004 (2024).

[32] J.-H. Song, J. van de Groep, S. J. Kim, M. L. Brongersma, Non-local metasurfaces for spectrally decoupled wavefront manipulation and eye tracking, Nat. Nanotechnol. 16, 1224-1230 (2021).

[33] M. Kim, D. Lee, J. Kim, J. Rho, Nonlocal metasurfaces-enabled analog light localization for imaging and lithography, Laser Photonics Rev., 18(7), 2300718 (2024).

[34] S. Droulias, T. Koschny, C. M. Soukoulis, Finite-size effects in metasurface lasers based on resonant dark states, ACS Photonics, 5(9), 3788-3793 (2018).

[35] L. M. Manez-Espina, A. Dıaz-Rubio, Control of the local and non-local electromagnetic response in all-dielectric reconfigurable metasurfaces, arXiv:2405.20201.

[36] Y. Xu, Y. Wang, Q. Zhou, L. Gao, Y. Fu, Unidirectional manipulation of Smith-Purcell radiation by phase-gradient metasurfaces, Opt. Lett., 48(15), 4133-4136 (2023).

[37] Y. Cao, Y. Fu, L. Gao, H. Chen, Y. Xu, Parity-protected anomalous diffraction in optical phase gradient metasurfaces, Phys. Rev. A, 107(1), 013509 (2023).

[38] M. Decker, I. Staude, M. Falkner, J. Dominguez, D. N. Neshev, I. Brener, T. Pertsch, Y. S. Kivshar, High-efficiency dielectric Huygens' surfaces, Adv. Optical Mater., 3(6), 813-820 (2015).

[39] Y. Xu, Y. Yang, D.-X. Yan, H. Duan, G. Zhao, Y. Liu, Gradient structure design of flexible waterborne polyurethane conductive films for ultraefficient electromagnetic shielding with low reflection characteristic, ACS Appl. Mater. Interfaces, 10(22), 19143-19152 (2018).

[40] L. V. R. Marcos, J. I. Larruquert, J. A. Méndez, J. A. Aznárez, Self-consistent optical constants of SiO2 and Ta2O5 films, Opt. Mater. Express, 6(11), 3622-3637 (2016).

[41] S. I. Azzam, A. V. Kildishev, Photonic bound states in the continuum: from basics to applications, Adv. Optical Mater. 9(1), 2001469 (2020).

[42] T. Barwicz, M. R. Watts, M. A. Popović, P. T. Rakich, L. Socci, F. X. Kärtner, E. P. Ippen, H. I. Smith, Polarization-transparent microphotonic devices in the strong confinement limit, Nature Photon. 1, 57-60 (2007).

[43] D. Dai, J. Bauters, J. Bowers, Passive technologies for future large-scale photonic integrated circuits on silicon:

polarization handling, light non-reciprocity and loss reduction, Light Sci. Appl., 1, e1 (2012).

[44] K. Koshelev, S. Kruk, E. Melik-Gaykazyan, J.-H. Choi, A. Bogdanov, H.-G. Park, Y. Kivshar, Subwavelength dielectric resonators for nonlinear nanophotonics, Science, 367 (6475), 288-292 (2020).

[45] Z. Ding, H. Li, X. Li, X. Fan, J. Jaramillo-Fernandez, L. Pattelli, J. Zhao, S. Niu, Y. Li, H. Xu, Designer $SiO_2$ metasurfaces for efficient passive radiative cooling, Adv. Mater. Interfaces, 11(3), 2300603 (2024).

[46] E. D. Mohamed Isa, H. Ahmad, M. B. Abdul Rahman, M. R. Gill, Progress in mesoporous silica nanoparticles as drug delivery agents for cancer treatment, Pharmaceutics, 13(2),152 (2021).

[47] H. Rai, K. RB Singh, S. S. Pandey, A. Natarajan, Role of Si and $SiO_2$ in optoelectronic device fabrication, J. Mol. Struct., 1316 (15), 138994 (2024).

[48] G. Zhu, D. Chao, W. Xu, M. Wu, H. Zhang, Microscale silicon-based anodes: Fundamental understanding and industrial prospects for practical high-energy lithium-ion batteries, ACS Nano, 15 (10), 15567-15593 (2021).

[49] C. F. Bohren and D. R. Huffman, Absorption and Scattering of Light by Small Particles (Wiley, New York, 1983).

[50] Z. C. Ruan, S. H. Fan, Superscattering of light from subwavelength nanostructures, Phys. Rev. Lett. 105(1), 013901 (2010).

[51] A. C. Valero, H. K. Shamkhi, A. S. Kupriianov, T. Weiss, A. A. Pavlov, D. Redka, V. Bobrovs, Y. Kivshar, A. S. Shalin, Superscattering emerging from the physics of bound states in the continuum, Nat. Commun. 14, 4689 (2023).

[52] W. Liu, Generalized Magnetic Mirrors, Phys. Rev. Lett. 119(12), 123902 (2017).

[53] W. Liu, Superscattering pattern shaping for radially anisotropic nanowires, Phys. Rev. A, 96(2), 023854 (2017).

[54] COMSOL, "Multiphysics@ v. 5.2. cn.comsol.com. COMSOL AB, Stockholm, Sweden,".

[55] Z. Ma, S. M. Hanham, Y. Gong, M. Hong, All-dielectric reflective half-wave plate metasurface based on the anisotropic excitation of electric and magnetic dipole resonances, Opt. Lett. 43(4), 911-914 (2018).

[56] R. Peng, Q. Zhao, Y. Meng, S. Wen, J. Zhou, Switchable all-dielectric magnetic-electric mirror based on higher-order dipoles, Phys. Rev. Applied, 13 (5), 054031 (2020).

[57] Hao Song, Lei Sun, Guo Ping Wang, Tunable perfect magnetic mirrors and retroreflectors in terahertz band, Opt. Express 28 (1), 753-759 (2020).

[58] B. Munkhbat, P. Wróbel, T. J. Antosiewicz, T. O. Shegai, Optical Constants of Several Multilayer Transition Metal Dichalcogenides Measured by Spectroscopic Ellipsometry in the 300-1700 nm Range: High Index, Anisotropy, and Hyperbolicity, ACS Photonics, 9(7), 2398-2407 (2022).

[59] H. Chen, L. Gao, Anomalous electromagnetic scattering from radially anisotropic nanowires, Phys. Rev. A, 86(3), 033825 (2012).

[60] W. Liu, Y. S. Kivshar, Generalized Kerker effects in nanophotonics and meta-optics, Opt. Express, 26(10), 13085-13105 (2018).

[61] H. Song, B. Hong, N. Wang, G. P. Wang, Kerker-type positional disorder immune metasurfaces, Opt. Express 31(15), 24243-24259 (2023).

[62] J. Olmos-Trigo, C. Sanz-Fernández, D. R. Abujetas, J. Lasa-Alonso, N. de Sousa, A. García-Etxarri, J. A. Sánchez-Gil, G. Molina-Terraza, and J. J. Sáenz, Kerker conditions upon lossless, absorption, and optical gain regimes, Phys. Rev. Lett. 125(7), 073205 (2020).

[63] X. Yin, J. Jin, M. Soljačić, C. Peng, B. Zhen, Observation of topologically enabled unidirectional guided resonances, Nature, 580, 467-471 (2020).

[64] H. Song, X. Zhang, J. Wang, Y. Sun, G. P. Wang, Bound state in the continuum and polarization-insensitive electric mirror in a low-contrast metasurface, Opt. Express 32(15), 26867-26883 (2024).

[65] J. D. Joannopoulos, S. G. Johnson, J. N. Winn, and R. D. Meade, Photonic Crystals: Molding the Flow of Light

-Second Edition (REV-Revised, 2), (Princeton University Press, 2008).

[66] J. D. Jackson, Classical electrodynamics, Wiley, (1999).

[67] W. Liu, A. E. Miroshnichenko, Beam Steering with Dielectric Metalattices, ACS Photonics, 5 (5), 1733-1741 (2018).

[68] A. Palmieri, A. H. Dorrah, J. Yang, J. Oh, P. Dainese, F. Capasso, Do dielectric bilayer metasurfaces behave as a stack of decoupled single-layer metasurfaces?, Opt. Express 32 (5), 8146-8159 (2024).

[69] R. Venkitakrishnan, Y. Augenstein, B. Zerulla, F. Z Goffi, M. Plum, C. Rockstuhl, On the physical significance of non-local material parameters in optical metamaterials, New J. Phys., 25 (12), (2023) 123014.

[70] H. Song, B. Hong, Y. Qiu, K. Yu, J. Pei, G. P. Wang, Tunable bilayer dielectric metasurface via stacking magnetic mirrors, Opt. Express, 30(13), 22885-22900 (2022).

[71] R. Y. Wu, S. He, J. W. Wu, L. Bao, T. J. Cui, Multi-frequency amplitude-programmable metasurface for multi-channel electromagnetic controls, Nanophotonics, 12 (13), 2433-2442 (2023).

[72] J. C. Lian, L. Zhang, Z. Luo, R. Z. Jiang, Z. W. Cheng, S. R. Wang, M. K. Sun, S. Jin, Q. Cheng, T. J. Cui, A filtering reconfigurable intelligent surface for interference-free wireless communications, Nat Commun., 15, 3838 (2024).

**Figures**

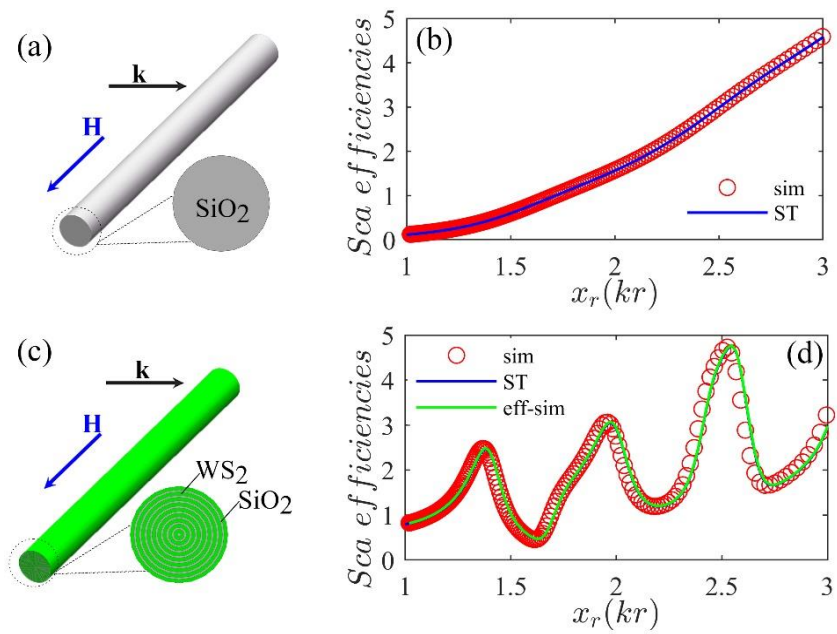

Figure 1. Scattering of an infinite-long cylinder. (a) Schematic of a SiO$_2$ cylinder normally illuminated by a TE plane wave with wavevector ***k*** and magnetic field **H**. (b) Total scattering efficiencies of the cylinder. Size parameter $x_r$ equals wavenumber $k$ times radius $r$. Red circles denote the results of numerical simulation (sim), and the blue curve was calculated by the scattering theory (ST). (c) configuration of a single radial anisotropic multiple-layer (ML) cylinder under the same TE wave. (d) Total scattering efficiencies of the ML cylinder. The green curve (eff-sim) is the result of numerical simulation with the effective material parameters.

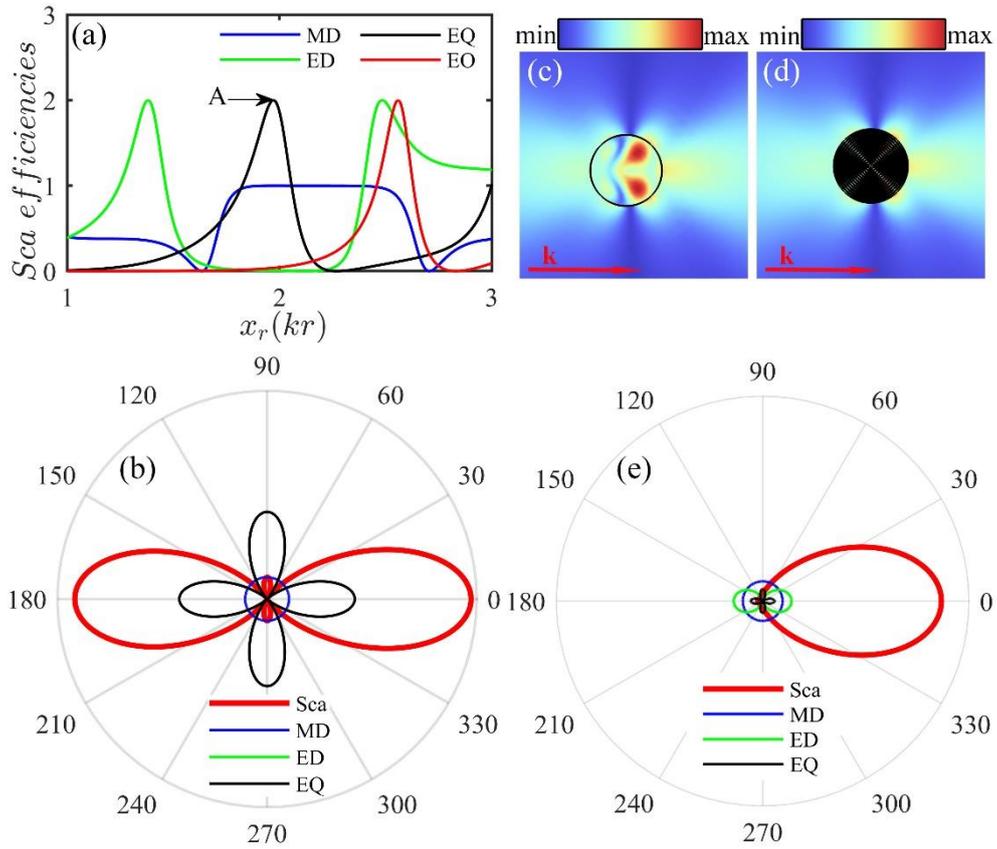

Figure 2. (a) The multipolar scattering efficiencies of the ML cylinder via the scattering theory. Point A denotes the $x_r$=1.971. (b) Angular scattering of the ML cylinder at point A. (c) Far-field scattering electric field |$\mathbf{E_s}$| at point A obtained by the numerical simulation with effective material parameters. $\mathbf{k}$ is the incident wavevector. (d) |$\mathbf{E_s}$| of the actual ML cylinder at point A. (e) Angular scattering of the SiO$_2$ cylinder at point A.

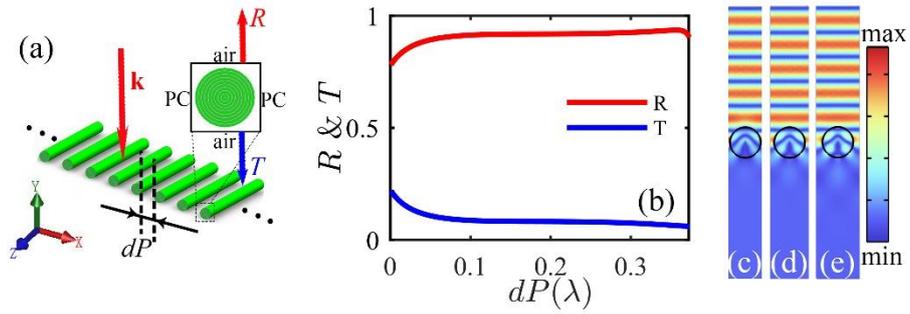

Figure 3. Reflectivity (*R*) and transmissivity (*T*) responses of infinite-size local metasurface under the TE wave. (a) The geometry of the metasurface is composed of the ML cylinders. Black solid frame indicates the schematic of one unit cell in numerical simulation. (b) *R* and *T* of the metasurface with effective material parameters. The minimal surface spacing between adjacent cylinders is *dP*, normalized by incident wavelength. (c)-(e) Electric field |$E_x$| distributions of unit cell with *dP*=0.1, 0.2, 0.3, respectively.

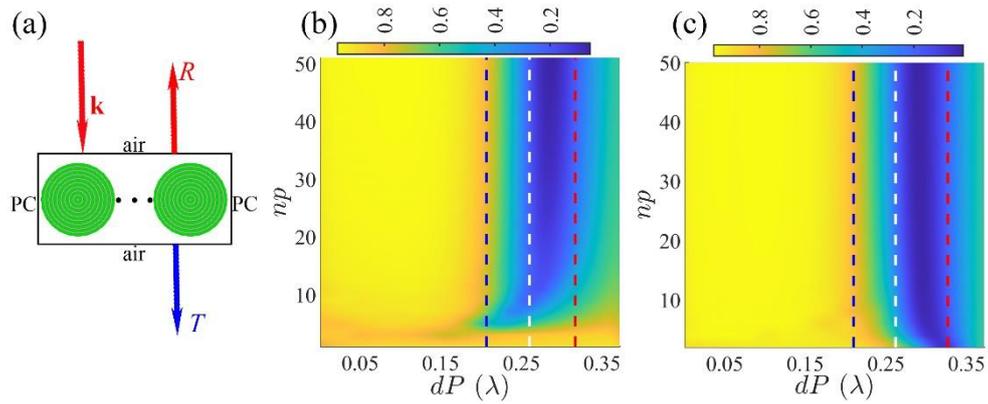

Figure 4. *R* and *T* of the infinite-size non-local metasurface illuminated by the TE wave. (a) Schematic diagram of the supercell. (b) *R* as a function of odd *np* and *dP*. *np* is the particle number in the supercell. Four regions are divided by the blue, white, and red dash lines. (c) *R* as functions of even *np* and *dP*.

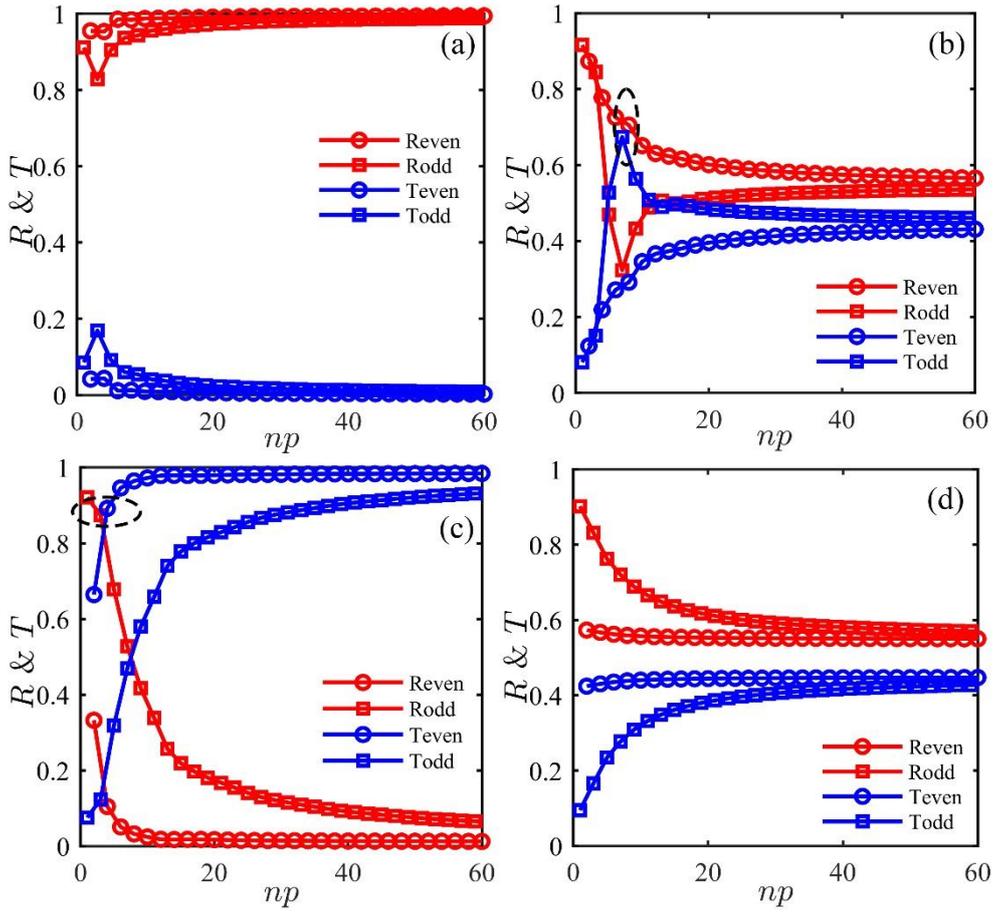

Figure 5. *R* and *T* spectra as functions of *np* at points B-E. (a)-(d) *R* and *T* spectra at points B, C, D, and E, respectively. Reven and Teven denote the *R* and *T* of the even *np*, respectively. Rodd and Todd indicate the cases of the odd *np*.

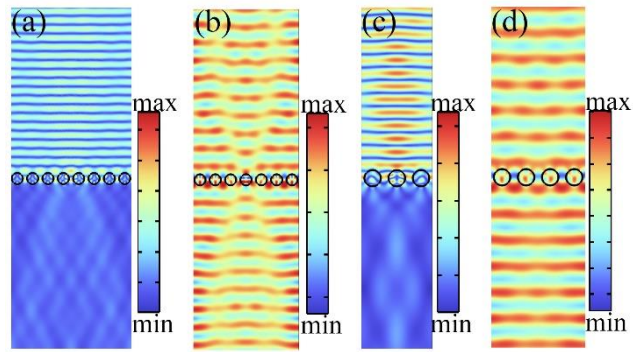

Figure 6. Reversal between equal magnitude high reflection and high transmission with adjacent *np*. (a) |$E_x$| of *np*=8 at point C. (b) $E_x$ of *np*=7 at point C. (c) |$E_x$| of *np*=3 at point D. (d) $E_x$ of *np*=4 at point D.

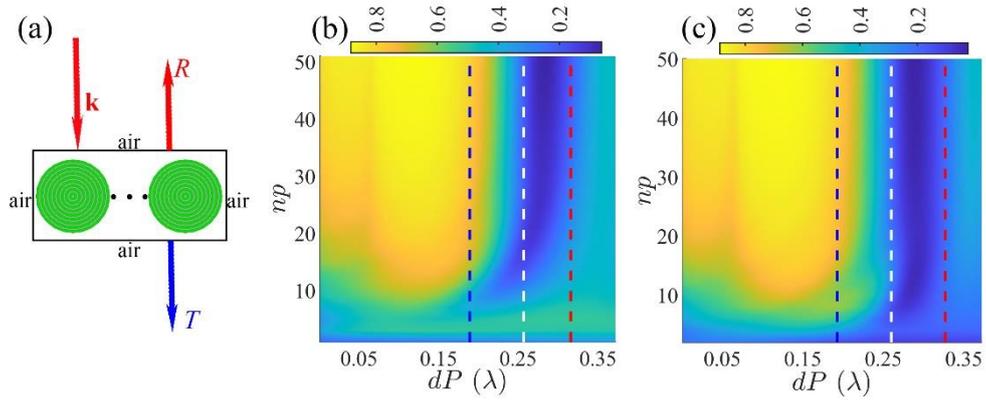

Figure 7. *R* and *T* of the finite-size non-local metasurface illuminated by the TE wave. (a) Schematic diagram of the ML cylinder arrangement in the metasurface. (b) *R* as a function of odd *np* and *dP*. (c) *R* as a function of even *np* and *dP*.

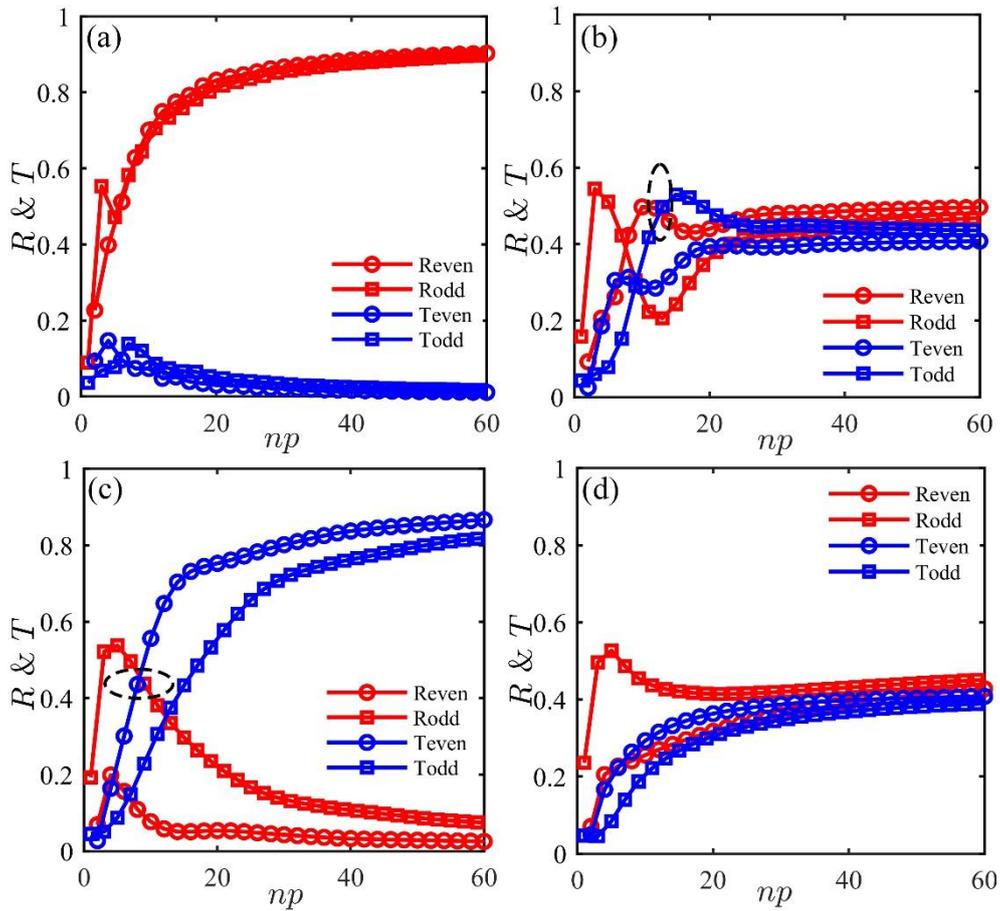

Figure 8. (a)-(d) *R* and *T* spectra of the finite-size non-local metasurface at points B, C, D, and E, respectively.

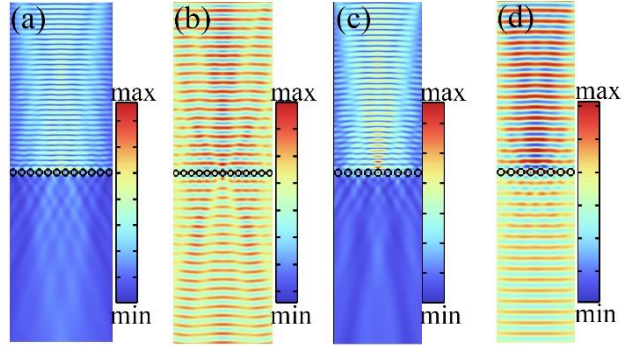

Figure 9. Reversal between equal magnitude high reflection and high transmission in the finite-size metasurface with adjacent *np*. (a) |$\mathbf{E}_x$| of *np*=12 at point C. (b) $\mathbf{E}_x$ of *np*=13 at point C. (c) |$\mathbf{E}_x$| of *np*=9 at point D. (d) $\mathbf{E}_x$ of *np*=8 at point D.

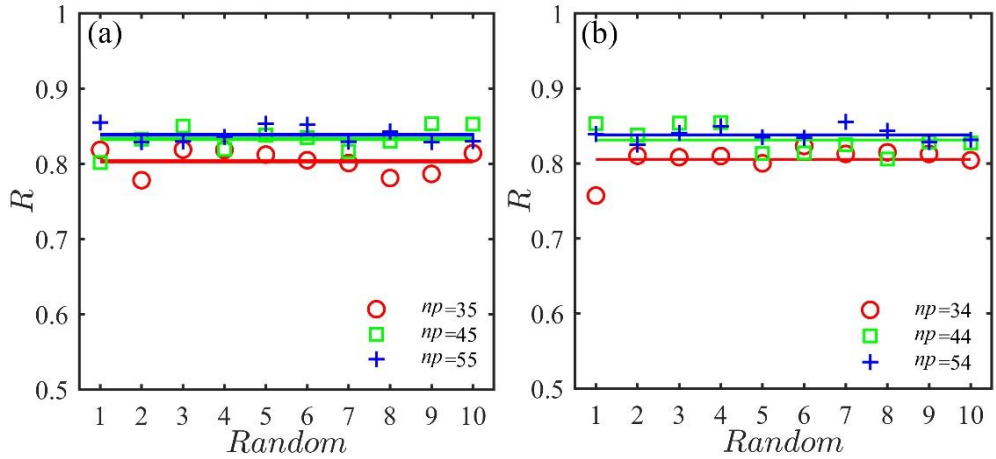

Figure 10. Disorder immunity of high reflection finite-size metasurfaces with randomly arranged cylinders. *dP* is random in the range of 0.01 to 0.183. (a) *R* of the random metasurfaces with odd *np*. *np*=35 is denoted by a red circle. The red line is the average *R* of the 10 samples. The green square and blue cross are *np*=45 and 55, respectively, thus green and blue lines also are their average values correspondingly. (b) *R* of the even *np* cases. Red circle, green square, and blue cross for the cases of *np*=34, 44, and 54, respectively. Similarly, the lines are corresponding average values.

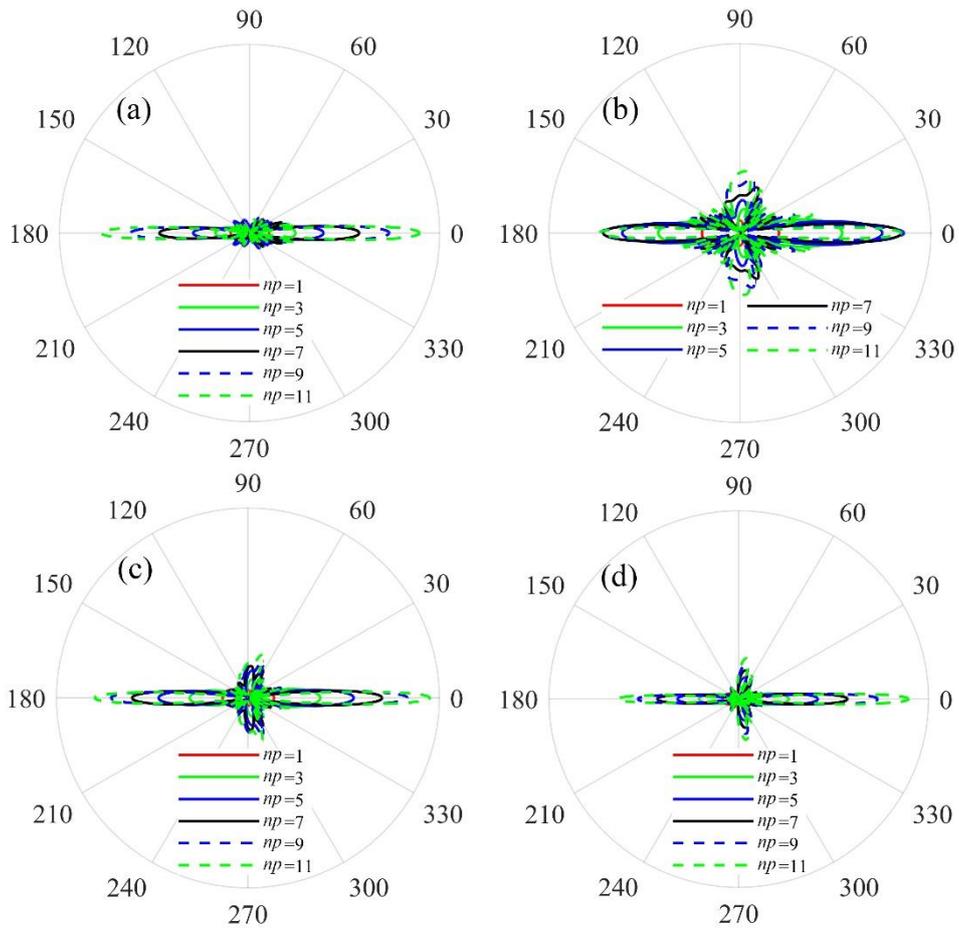

Figure 11. Angular multiple scatterings of odd *np*. Multiple scatterings (a) at point B, (b) at point C, (c) at point D, and (d) at point E.

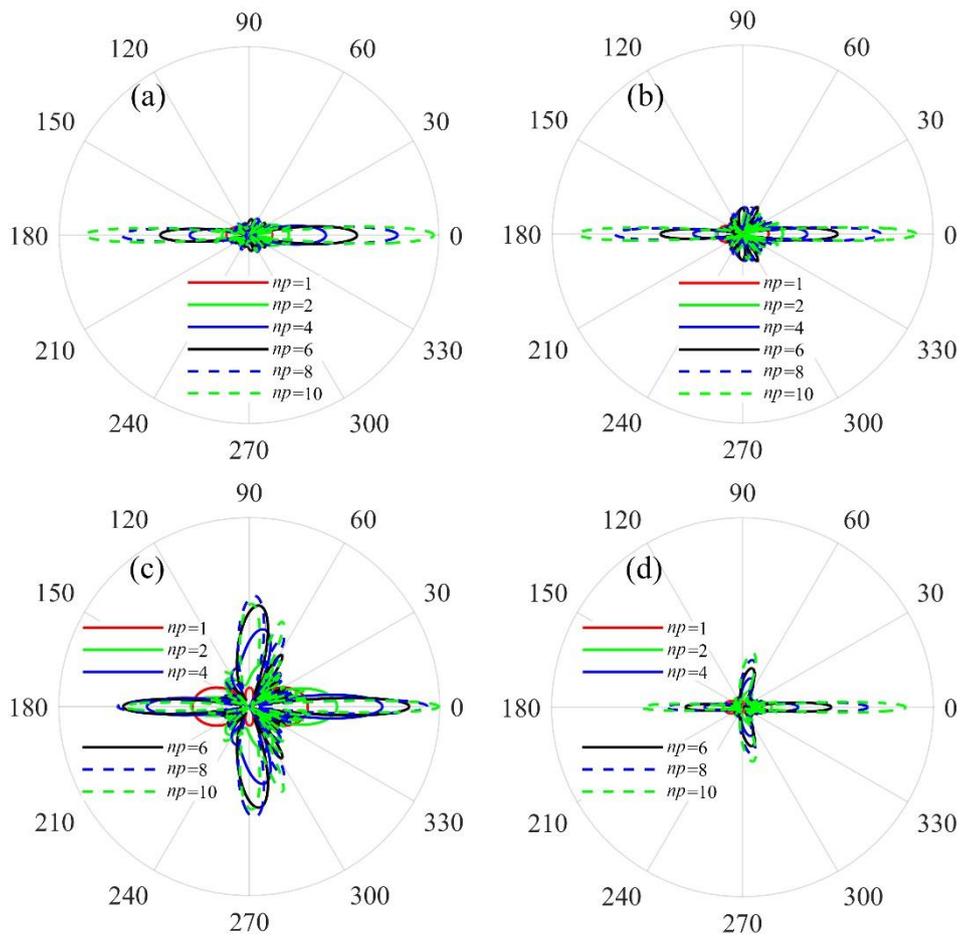

Figure 12. Angular multiple scatterings of even *np*. Multiple scatterings (a) at point B, (b) at point C, (c) at point D, and (d) at point E.

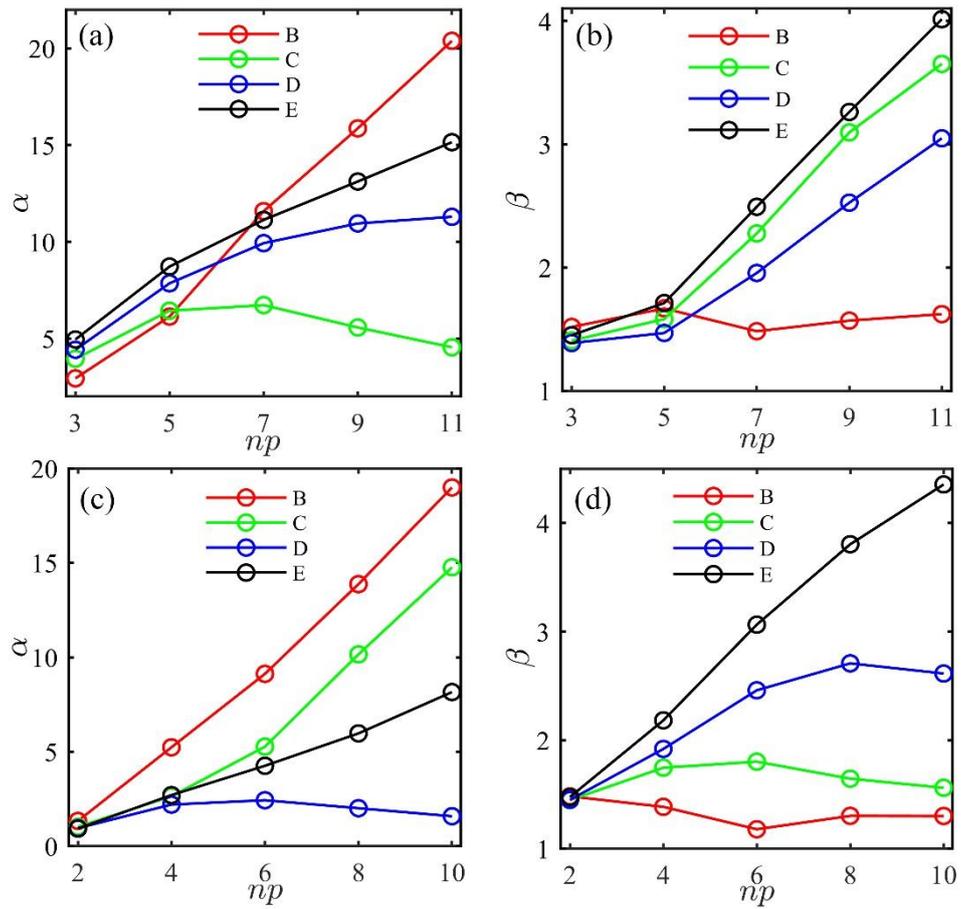

Figure 13. $\alpha$ and $\beta$ as functions of $np$. $\alpha$ is the backward multiple scattering intensity normalized by the single-cylinder case. $\beta$ is the ratio of lateral multiple scattering intensity normalized by the case of a single cylinder. (a) and (b) for the $\alpha$ and $\beta$ of odd $np$, respectively. (c) and (d) for $\alpha$ and $\beta$ of even $np$, respectively.